\documentclass[
    journal,
    ]{IEEEtran}
\usepackage{graphicx}
\usepackage{amsmath}
\usepackage{amssymb}
\usepackage{afterpage}
\usepackage{verbatim}
\usepackage{psfrag}
\usepackage{color}
\usepackage{cite}
\usepackage{setspace}
\usepackage{epsfig}
\usepackage{epstopdf}
\usepackage{footmisc}
\usepackage{fmtcount}
\usepackage{stackrel}
\usepackage{float}
\usepackage{breqn}
\usepackage{enumerate}
\usepackage{graphicx}
\usepackage{subcaption}
\usepackage{hyperref}
\usepackage{hypcap}
\usepackage[ruled,vlined,linesnumbered]{algorithm2e}
\usepackage[normalem]{ulem}
\usepackage{authblk}

\makeatletter 
\newcommand\semiHuge{\@setfontsize\semiHuge{23}{28}}
\makeatother

\begin{document}

\title{On Distributed Routing in Underwater Optical Wireless Sensor Networks}
\author[1]{Rawan Alghamdi }
\affil[1]{Department of Electrical and Computer Engineering, Effat University, Jeddah, 22332, Saudi Arabia. }
\author[2]{Nasir Saeed }
\affil[2]{Computer, Electrical, and Mathematical Sciences \& Engineering (CEMSE) Division, King Abdullah University of Science and Technology (KAUST), Thuwal,  23955-6900, Saudi Arabia. }
\author[1]{ Hayssam Dahrouj }
\author[2]{Tareq Y. Al-Naffouri }
\author[2]{and Mohamed-Slim Alouini }

\maketitle
\thispagestyle{empty}
\pagestyle{empty}

\begin{abstract}
Underwater optical wireless communication (UOWC) is becoming an attractive technology for underwater wireless sensor networks (UWSNs)  since it offers high-speed communication links. Although UOWC overcomes the drawbacks of acoustic and radio frequency communication channels  such as high latency and low data rate, yet, it has its own limitations. One of the major limitations of UOWC is its limited transmission range which demands to develop a multi-hop network with efficient routing protocols. Currently, the routing protocols for UOWSNs are centralized having high complexity and large end-to-end delay. In this article, first, we present the existing routing protocols for UOWSNs. Based on the existing protocols, we then propose distributed routing protocols to address the problems of high complexity and large end-to-end delay. Numerical results have been provided to show that the proposed routing  protocol is superior to the existing protocols in terms of complexity and end-to-end delay. Finally, we have presented open research directions in UOWSNs.

\end{abstract}
\IEEEpeerreviewmaketitle


\section{Introduction}
The survival of Planet Earth depends on the Oceans. The Oceans on Earth surface flow for more than three-quarters having 97 \% of the water which produces an enormous amount of Oxygen that we breathe and absorb the Carbon. Indeed, there are so many benefits that Oceans provide to the humankind such as Oxygen, food, climate regulation, recreation, transportation, and medicine.    It is reported in \cite{noaa} that businesses based on Oceans contribute more than 500 billion US dollars to the world economy. Despite all these benefits, 95 \% volume of the Oceans is unexplored. 
\begin{figure}
\centering
\includegraphics[width=0.9\columnwidth]{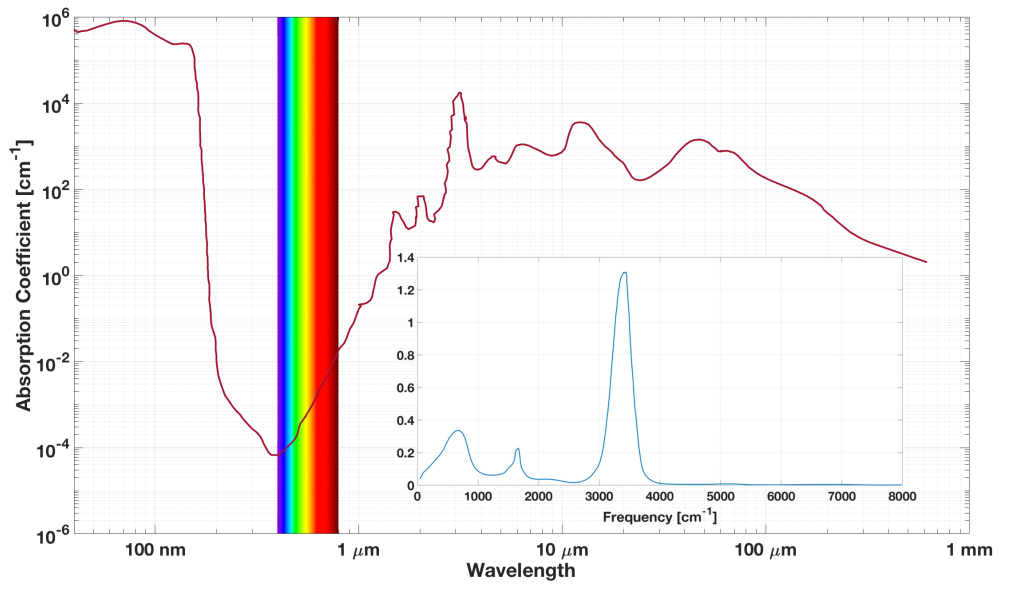}  
\caption{Attenuation of optical waves in the aquatic medium \cite{survey}.\label{fig:attenuation}}  
\vspace{-1.5 em}
\end{figure}
The unexplored regions of Oceans can be investigated by using underwater wireless sensor networks (UWSNs). UWSNs have the great potential to enable numerous applications such as offshore and ocean sampling, ocean exploration, navigation, recreation, surveillance, and disaster management. In order to meet the demands of such applications, an underwater wireless communication technology with high data rate and transmission speed becomes essential to establish a real-time broadband communication.  Efforts have been made by the researchers to provide a variety of underwater wireless communication systems by using different types of waves.
Distinct types of waves propagate underwater, e.g., electromagnetic waves in the radio frequency (RF) band, acoustic waves, and optical waves in the visible spectrum. The electromagnetic waves in RF band are highly attenuative in the seawater, bounded by the extremely low-frequency band and limited to shallow waters; therefore, it supports insufficient and limited data rate communications \cite{book}. Additionally, acoustic waves which are mostly utilized for different underwater applications support long communication ranges in underwater environments up to tens of Km, yet it suffers from high latency due to the low transmission speed. Due to the low transmission speed of acoustic waves in underwater, it cannot achieve high data rate performance \cite{Arnon, book}.
\begin{figure*}
\centering
\includegraphics[width=1.3\columnwidth]{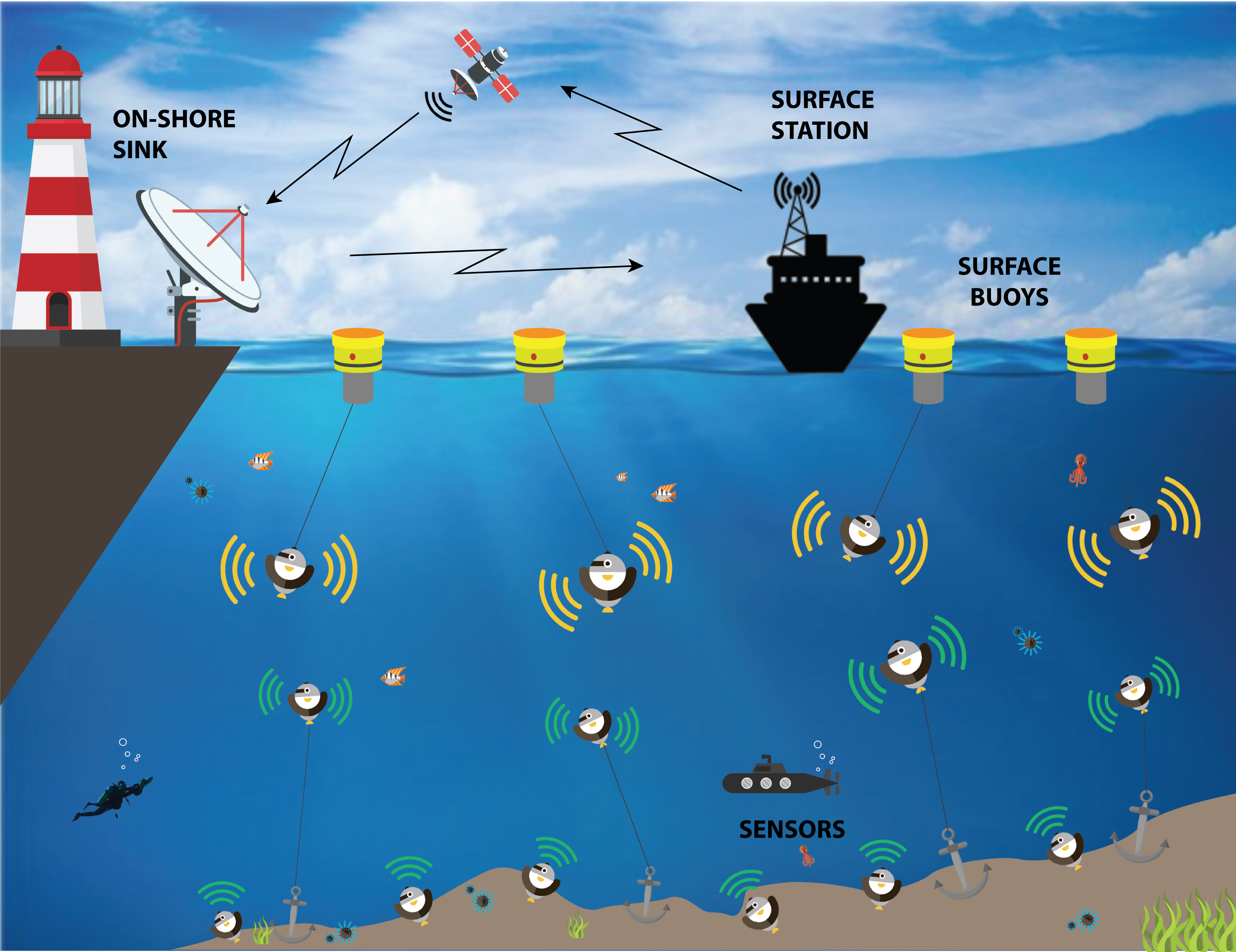}  
\caption{An architecture of underwater optical wireless sensor network.\label{fig:systemmodel}}
\hrule  
\end{figure*}

On the contrary,  underwater optical wireless communication (UOWC) is considered as an alternative approach for providing high-speed transmission and data rate. The advantages of UOWC include supporting higher bandwidth communication and providing broadband links even in the presence of high absorption and scattering \cite{Arnon}. UOWC provides high propagation speed almost as much as the speed of light with low latencies at the cost of short communication ranges up to tens of meters.  UOWC also has the advantage of having small size transceiver devices, low installation, and operational cost. Unfortunately, the physical properties of Seawater and intrinsic properties of light limit the performance and reliability of UOWC systems \cite{book}.  Although optical waves are highly absorbed and scattered in the water, the visible spectrum and especially the blue/green region has low absorption in Seawater as shown in Fig.~\ref{fig:attenuation} \cite{survey}. Therefore, utilizing a wavelength within the blue/green region guarantees high-speed underwater wireless communication.  Hence, a network of sensors is formed which can communicate by using UOWC referred as an underwater optical wireless sensor network (UOWSN).  Above all, due to the limitations of UOWSNs such as limited transmission range, low transmission power, and intrinsic properties of light in water it is required to develop a multi-hop network where the sensing information is transmitted over a multi-hop path or series of relaying nodes. Using multi-hop routing has the advantage of increasing the network lifetime, coverage, extending the communication range, and reducing the interference level \cite{ABER}.

Undoubtedly, routing protocols for multi-hop networks are essential and are considered as one of the critical design elements to achieve a reliable and efficient communication link. Routing protocols in simple definition determine that how information travels from a source node to the target node. Additionally, an efficient and reliable routing protocol plays a significant role in multi-hop networks since it is responsible for choosing a path between the source and the target with optimal performance based on the design requirements \cite{Routing}.

Based on the decision making process routing protocols can be classified into two categories, centralized and distributed (decentralized). In centralized routing protocols, a central node is responsible  for collecting all the information from the network and then estimate the best path from source to target for a given performance metric. Centralized routing uses a sophisticated algorithm to choose the optimal path which satisfies the design requirements. Centralized routing protocols commonly utilize Dijkstra's algorithm to estimate the shortest path between the source and the target.   However, in dense sensor networks, collecting all the information from the network and then computing all possible paths to select the optimal one is a cumbersome task which affects the performance of the protocol in terms of latency and complexity. 

On the other hand, the distributed routing protocols use a more straightforward process where a central node is not required. In fact, the source node decides on the forwarding node independently using the local information available. 
In other words, the selected path is updated by each forwarding node until the data packet reaches the target. One of the challenges faced by the distributed routing protocols is that the selection of forwarding nodes based on the design metric may not result in the optimal path. The literature on routing protocols for terrestrial networks and underwater acoustic wireless sensor networks (UAWSNs) is rich where a large number of surveys are presented on this topic. A large number of centralized and distributed routing protocols such as depth-controlled routing \cite{DCR}, focused beam routing \cite{FBR}, vector base routing \cite{VBF}, and sector base routing \cite{SBR} have been proposed in the past decade for UAWSNs. However, the routing protocols for both terrestrial networks and UAWSNs are not suitable for UOWSNs. Therefore, reliable and efficient routing protocols for UOWSNs needs to be investigated which considers the propagation characteristics of light in the underwater medium. Recently, a centralized routing protocol (CRP) is proposed in \cite{Nasir} for UOWSNs. However, the nature of the CRP proposed in \cite{Nasir} is centralized with large end-to-end time delay and high complexity. Henceforth, developing a routing protocol with an efficient end-to-end (E2E) path characterization is necessary for UOWSNs. The selected path varies with the variation of the evaluation metrics of the routing protocol used. According to \cite{Nasir} the multi-hop path is an optimal path if it provides the maximum data rate. However, the data rate is not the only evaluation metric to characterize the performance of an end to end path between the source and the target. There are other evaluation metrics to design routing protocols such as the nature of the protocol (centralized or distributed), complexity, end-to-end bit error rate (BER), and end-to-end time delay.

In this paper, we investigate different routing protocols for UOWSNs in terms of end-to-end BER, complexity, and end-to-end time delay. In specific the contributions of the paper are as follows:
\begin{itemize}
\item The characteristics of UOWC is examined in the line of sight (LoS) condition with different types of Seawater and different beam-widths of the optical signal.
\item Selection of the forwarding node is investigated based on the CRP \cite{Nasir}, the distributed routing protocol (DRP), and the proposed distributed sector based routing protocol (SRP).
\item Evaluation of each protocol in terms of the design parameters, i.e., end-to-end BER, complexity, and end-to-end time delay. The proposed SRP has low complexity and low end-to-end time delay.
\item The paper further provides several future research directions mainly underwater channel characterization, cross-layer optimization, advanced routing protocols,  hybrid underwater wireless communication, energy harvesting,  and real-life implementation of UOWSNs.
\end{itemize}

In the remainder of the paper, we first define the network architecture, calculate the link budget of UOWC channel, and present the bit error rate calculation. Then we present three routing protocols CRP, DRP, and SRP and evaluate their performance. Finally, we provide future research directions and conclude the paper.


\section{Underwater Optical Wireless Communication} 
\label{systmodel}
In this section, we first briefly introduce underwater propagation characteristics of optical waves and their limitations. Then we present the architecture of an UOWSN followed by the calculation of a link budget and BER for UOWSNs.
\subsection{Underwater Propagation Characteristics of Optical Waves}
\label{propgartion}	
Optical waves are considered as a promising transmission carrier for underwater wireless communication. Compared to the other types of waves for underwater wireless communication, optical waves can achieve broadband links with high data rate (Gbps) and low latency due to the high speed of light in water \cite{survey,book}. To understand the propagation characteristics of optical waves in the aquatic medium, the inherent and apparent properties of optical waves need to be discussed. The inherent properties of optical waves in underwater are highly dependent on the impairments of the underwater transmission medium such as absorption, scattering, salinity, and temperature \cite{survey, inherent}. Among inherent properties, the light attenuation can be primarily determined by the absorption and scattering coefficients. Absorption and scattering can be defined as the photon energy transferred into another form and the interaction of photon particles with the transmission medium's molecules and particles, respectively \cite{advances, book}. Absorption of the photon particles in the underwater transmission medium results in the reduction of communication range while scattering reduces the number of collected photons at the receiver (low SNR) \cite{Zeng2017}. The research work in \cite{inherent2, inherent3, inherent4, inherent5, inherent6, inherent7} has only considered the inherent properties of light to model the UOWC channel. The apparent properties of optical waves depend on both the underwater transmission medium and the geometrical structure of the light field such as radiance, irradiance, reflectance, and the light beam's parameters \cite{Zeng2017,apparent}. The authors in \cite{apparent2,apparent3, Analysis} take into account the apparent optical properties as a function of inherent properties to find a general relationship between them. Nevertheless, we refer the interested readers to \cite{survey,  Zeng2017} which covers the different aspects of UOWCs.

\subsection{Limitations}
Despite the  useful features of UOWC over other underwater communication methods, UOWC suffers from different impairments caused by the prorogation of light in water. The different impairments include absorption, scattering, geometrical losses, interference from sunlight, salinity, turbulence and air bubbles. These impairments limit the transmission range, introduce multi-path fading, and causes misalignment between the transmitter and the receiver for UOWC. Scattering of optical waves spreads the light photons in arbitrary directions such that not all of the beam is received by the receiver, which leads to a delay in receiving some portion of the beam. Additionally, the scattering and absorption cause water molecules and particles to interact with light photons resulting in multi-path fading \cite{fading}. Moreover, UOWC suffers from short-term disconnection caused by the misalignment of the transceivers due to the sea surface movements \cite{misalign, randmov}. In short, all these impairments summed up to reduce the performance of UOWC systems and substantially shortens the communication ranges. Therefore, to get the most out of the UOWC systems, the routing protocols must consider the above mentioned physical layer characteristics and limitations.

\begin{figure*}
    \centering
    \begin{subfigure}[b]{0.44\textwidth}
\includegraphics[width=1\columnwidth]{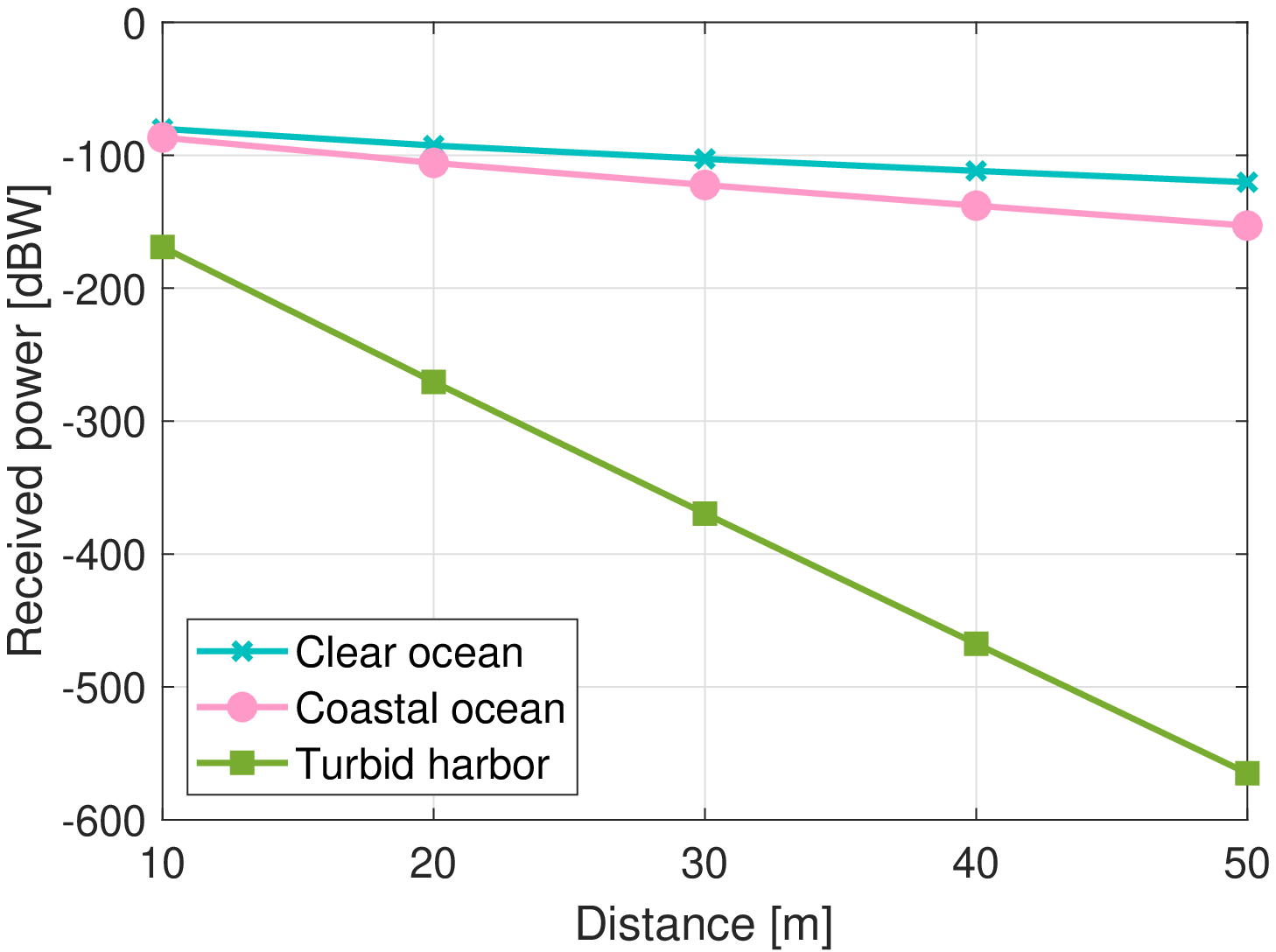}  
\caption{Recieved power Vs. types of water. \label{fig:Pr_3water}}   
    \end{subfigure}
    \begin{subfigure}[b]{0.44\textwidth}
\includegraphics[width=1\columnwidth]{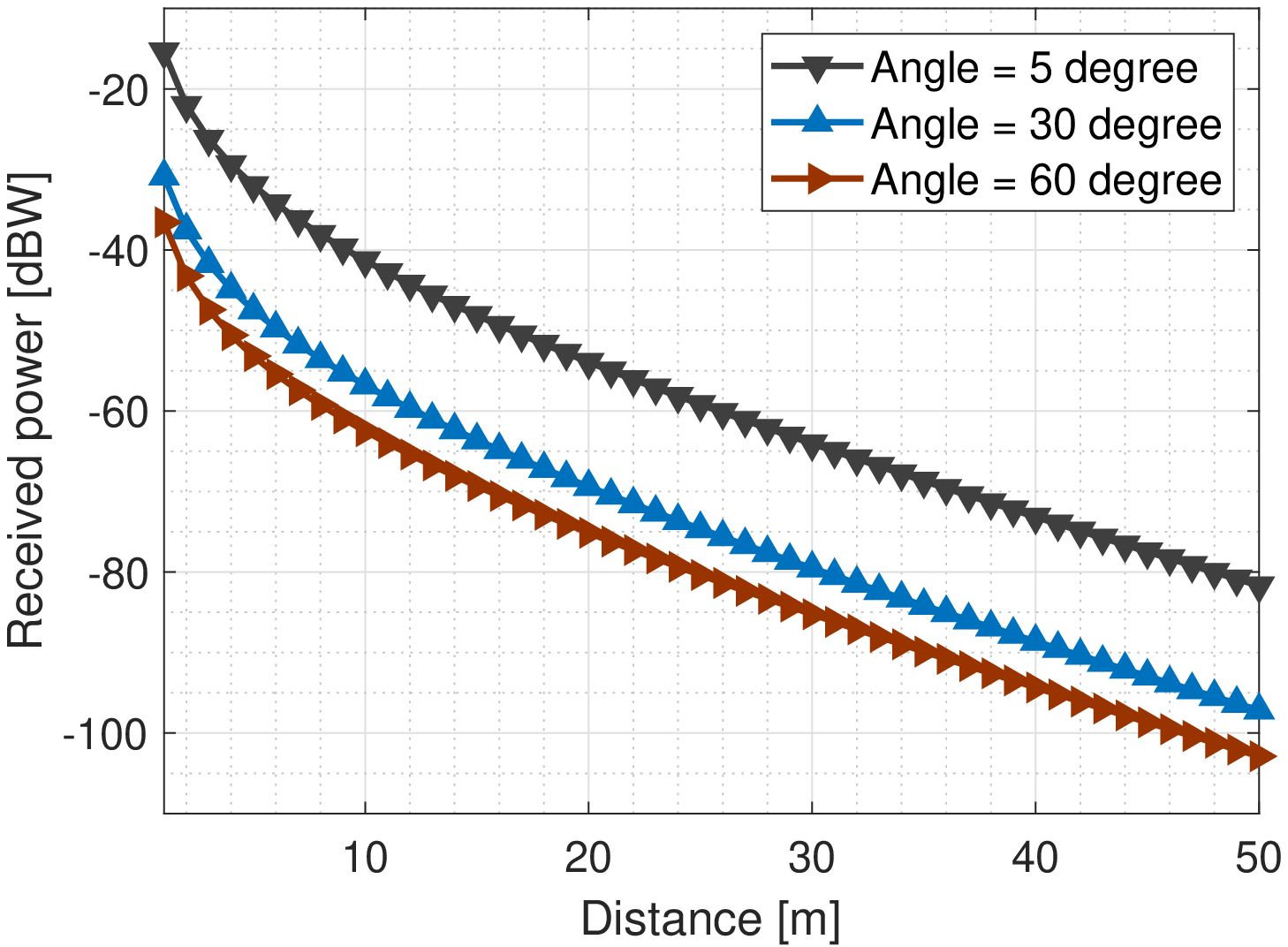}  
\caption{Recieved power Vs.  divergence angles. \label{fig:Pr_3angle}}  
    \end{subfigure}
    
    \begin{subfigure}[b]{0.44\textwidth}
\includegraphics[width=1\columnwidth]{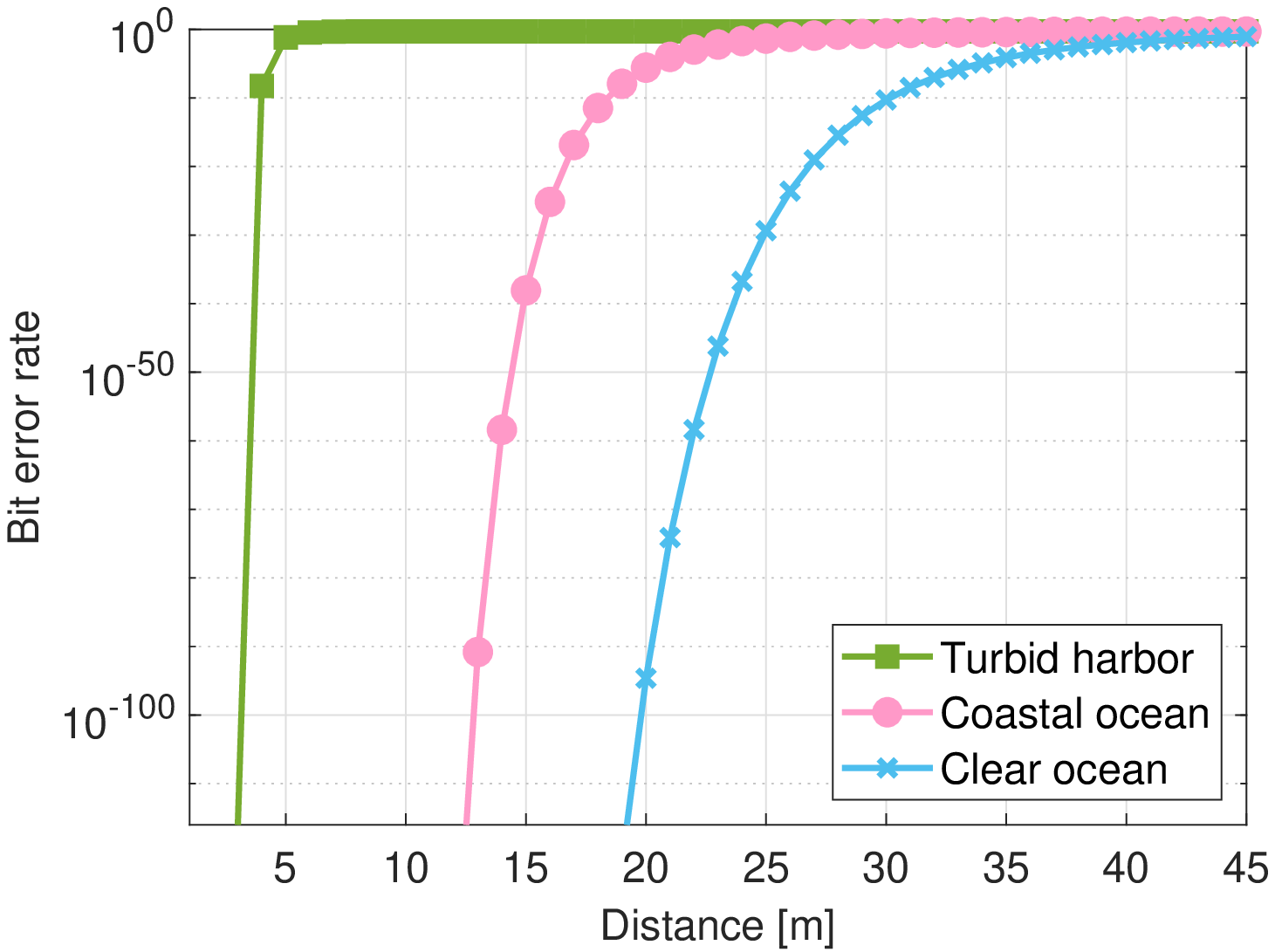}  
\caption{Bit error rate Vs. types of water.\label{fig:BER_3water}}  
    \end{subfigure}
    \begin{subfigure}[b]{0.44\textwidth}
\includegraphics[width=1\columnwidth]{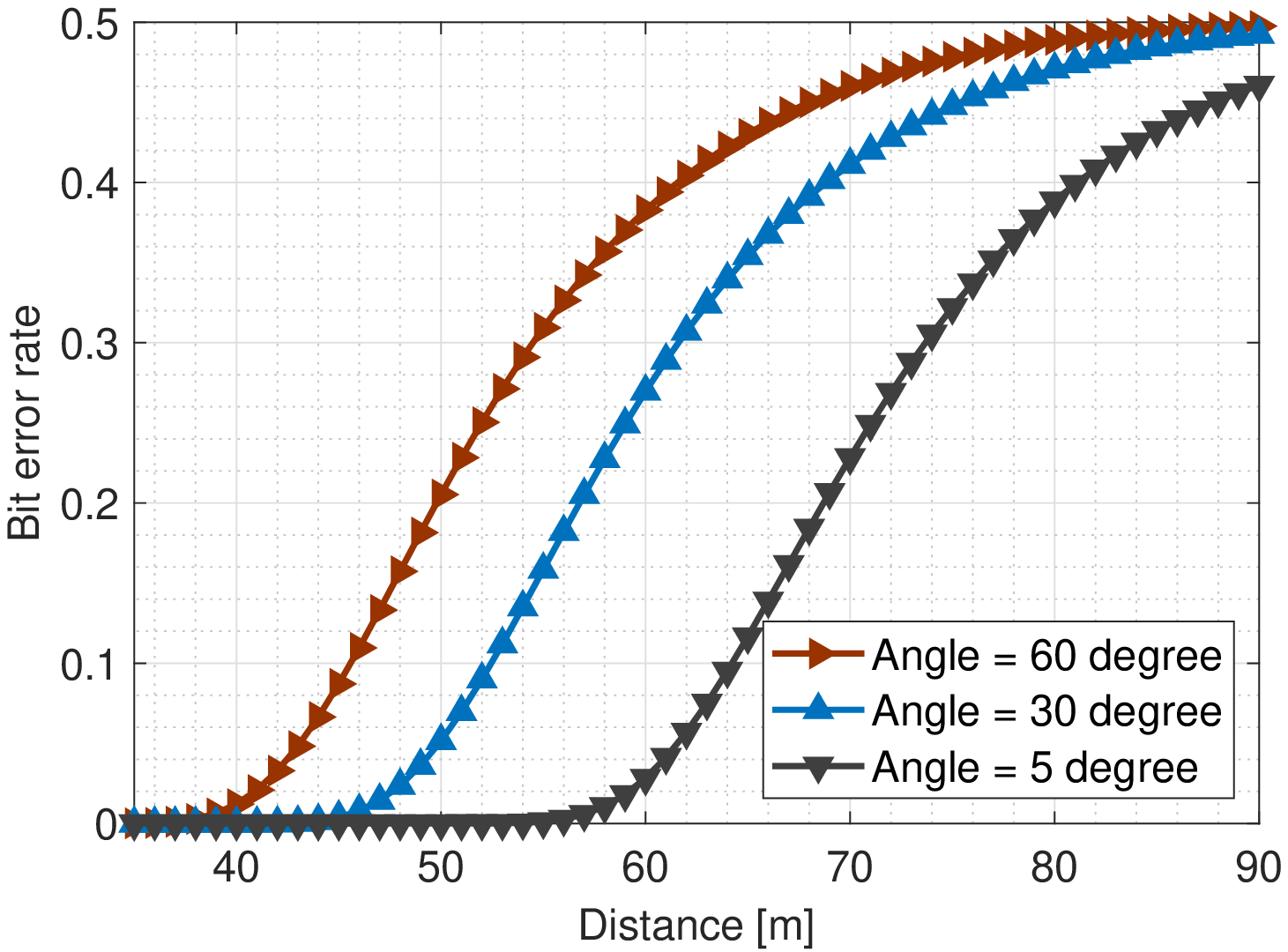}  
\caption{Bit error rate Vs. divergence angles. \label{fig:BER_3angle}}  
    \end{subfigure}    
\caption{Link budget and BER calculation of UOWC channel in LoS conditions.}
\label{fig:prop_connect}
\hrule
\end{figure*}

\subsection{Network Model}

In this paper, we consider an UOWSN which consists of $N=M+B$ number of sensor nodes, where $M$ is the number of sensor nodes which are either deployed at the seabed or moored, and $B$ is the number of surface buoys as shown in Fig.~\ref{fig:systemmodel}. The sensor nodes are responsible for monitoring its surroundings and send the collected information to the surface buoys in a  multi-hop fashion. The surface buoys are responsible for receiving the data collected from the sensors through UOWC link and disseminate the received information from the sensors to the surface station through a RF communication link. In the following, first, we show the simulation of the link budget for a UOWC link and then evaluate the single hop BER in LoS conditions. Finally, we extend the scenario of single hop BER calculation to the multi-hop case.

\subsubsection{UOWC Link Budget}
Optical waves propagating in the underwater environments are absorbed and scattered which is characterized by the Beer-Lambert' law \cite{beer}. The extinction coefficient for absorption and scattering is described as   
\begin{equation}
t(\lambda)= a(\lambda)+ s(\lambda),
\end{equation}
where $a(\lambda)$ represents the absorption coefficient, $s(\lambda)$ represents the scattering coefficient, and $\lambda$ is the wavelength \cite{book}. 
According to the extinction coefficient, the received power for a Line-of-Sight (LoS) link is described as follows \cite{Nasir}
\begin{equation}\label{eq:power}
P_{R_xlos}= P_{Tx} \eta_{Tx}\eta_{Rx} \exp\left(\frac{-t(\lambda)d}{\cos(\theta)}\right)\frac {A \cos(\theta)}{2\pi\left(1-\cos(\theta_0)\right)d^{2}},  
\end{equation} 
where $P_{Tx}$ is the average optical transmitter power, $\eta_T\eta_R$ are the  transmitter and receiver optical efficiencies, respectively. $\theta$ is the angle between the transmitter-receiver trajectory, $A$ is the  aperture area of the receiver, $\theta_0$ is the laser beam divergence angle, and $d$ is the Euclidean distance between the transmitter and the receiver. Based on the Beer-Lambert' law, the relationship of received power in LoS condition and different types of water (clear ocean, coastal ocean, and turbid harbor) is shown in Fig.~\ref{fig:Pr_3water} by using the parameters mentioned in Table \ref{table:para}. The received power increases with the decrease of the extinction coefficient $t(\lambda)$  when the transmission power and divergence  angle are kept constant which explains high received power in clear ocean water when compared to the other types of the water. Additionally, Fig.~\ref{fig:Pr_3angle} shows the impact of the laser beam width on the received power, where  narrowing the beamwidth results in receiving more power.

\subsubsection{Bit Error Rate}
Based on the received power from Beer-Lambert' law (\ref{eq:power}), and according to the central limit theorem, the bit error rate (BER) follows a Gaussian distribution. Therefore, using the BER is given in \cite{connectivity} as
\begin{equation}
P(e)= \frac{erfc\left[\sqrt{\frac{t}{2}}(\sqrt{r_1}-\sqrt{r_0})\right]}{2},
\end{equation}  
where $erfc(\cdot)$ is the complementary error function, $r_0$ and $r_1$ are the numbers of photon arrivals when transmitting a binary bit 0 and a binary bit 1, respectively. The number of photon arrivals for $r_1$ and $r_0$ is given as
\begin{equation}
r_0= r_{dc} +r_{bg},
\end{equation}  
\begin{equation}
r_1= r_{dc} +r_{bg}+r_p,
\end{equation}  
where $r_{dc}$ represents the source of additive noise due to dark counts, $r_{bg}$ is the source of additive noise due to background illumination, and $r_p$ is the photon arrival rate given by \cite{Arnon_2010} 
\begin{equation}
r_p= \frac{P_{R_x}\eta\lambda}{tRhc},
\end{equation}  
where $t$ is the pulse duration, $R$ is the data rate, $\eta$ is the detector counting efficiency, $h$ and $c$ are Plank's constant and the speed of light in the water, respectively.

The BER performance for different types of water is presented in Fig.~\ref{fig:BER_3water}, where BER performance is much better in clear ocean as compared to the coastal ocean and turbid harbor, due to the low extinction coefficient in clear ocean water. The extinction coefficient for clear water is less than both coastal ocean water and turbid harbor water. In addition, Fig.~\ref{fig:BER_3angle} represents the BER performance with different laser beam divergence angles, where the BER gets better in performance when the angle is narrow because the transmitted power is more focused towards the receiver.

\begin{figure}
\includegraphics[width=1\columnwidth]{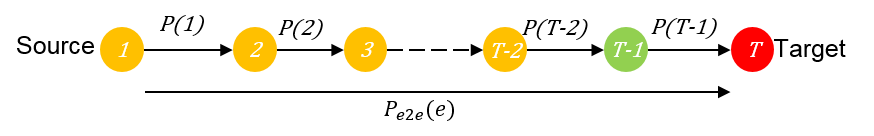}  
\caption{Multi-hop path with $T-1$ hops. \label{fig:E2EBER}}  
\end{figure}    

We calculate the end-to-end BER $P_{e2e}(e)$ in a multi-hop path by using the recursive joint probability in \cite{ABER} given as
\begin{equation}\label{eq:e2eber}
P_{e2e}(e) =(1-P(e)_{T-1})P(T)  + (1-P(T))P_{T-1}.
\end{equation}
where $P(i)$ is the single link BER from node $i$ to node $i+1$, $P_{i}$ is the end-to-end BER from source node to node $i+1$, and $i = \{1,2,3,...,T-1\}$. Therefore,  $(1-P_{i})$ is the probability of a bit transmitted correctly from a source node to the $i$-th node and $(1-P(i))$ is the probability of a  single link between node $i$ and node $i+1$.


 
%

\section{Routing Protocols for UOWSNs}
Routing plays a significant role in order to choose and sustain transmission paths in a communication system. Due to its significance, researchers have come up with various routing protocols, to maintain and advocate distinct channel characteristics. The limitations of UOWC necessitate developing an efficient routing protocol which takes into account the short communication ranges along with the propagation characteristics of optical waves in the aquatic medium. Routing protocols are optimized concerning different evaluation metrics such as end-to-end BER, delay efficiency, energy efficiency, and complexity.

In this section, first we introduce the centralized routing protocol as proposed in \cite{Nasir}, present a distributed version of the CRP, and then we propose a novel sectorized routing protocol  with low complexity and low end-to-end delay. Finally, we assess the performance of all the three routing protocols by providing numerical results.

\begin{figure}
    \centering
    \begin{subfigure}[b]{0.35\textwidth}
\includegraphics[width=1 \columnwidth]{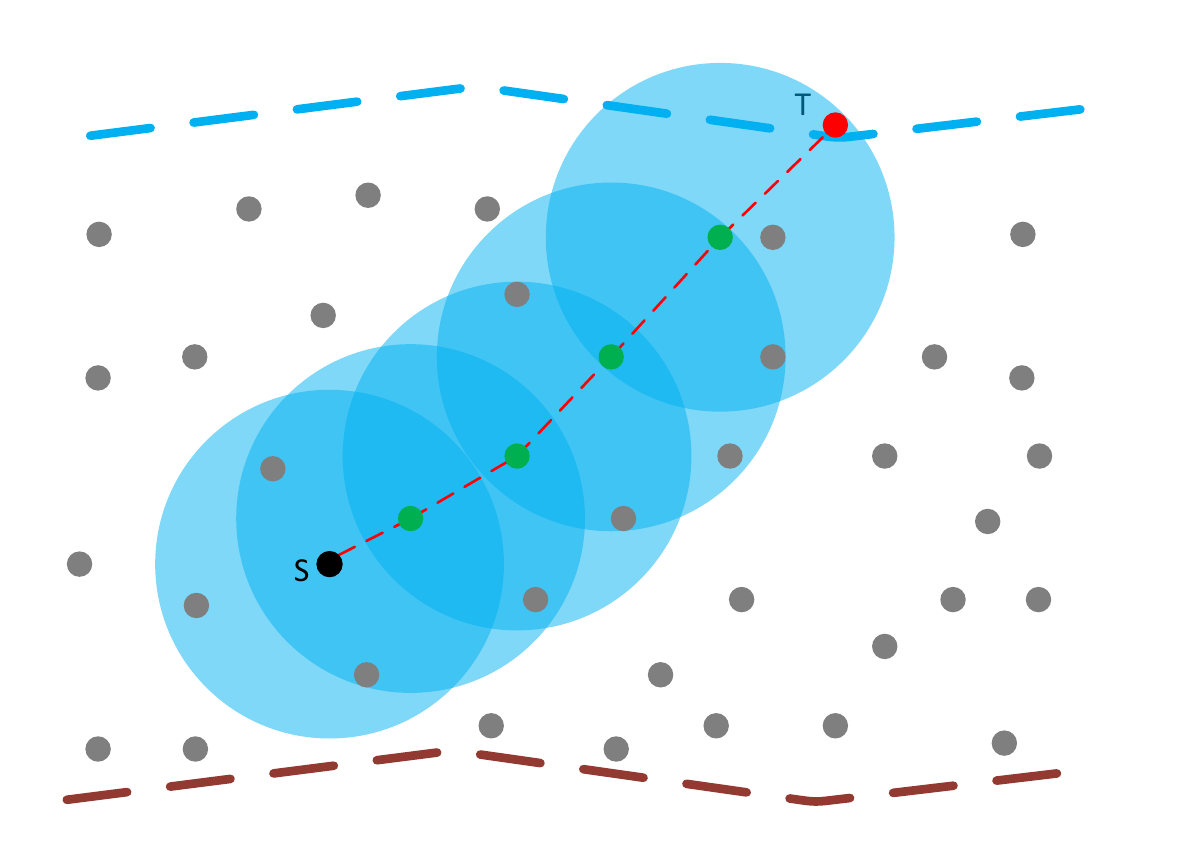}
    \caption{Centralized routing protocol (CRP).}
    \label{fig:crp}
    \end{subfigure}

    \begin{subfigure}[b]{0.35 \textwidth}
\includegraphics[width=1 \columnwidth]{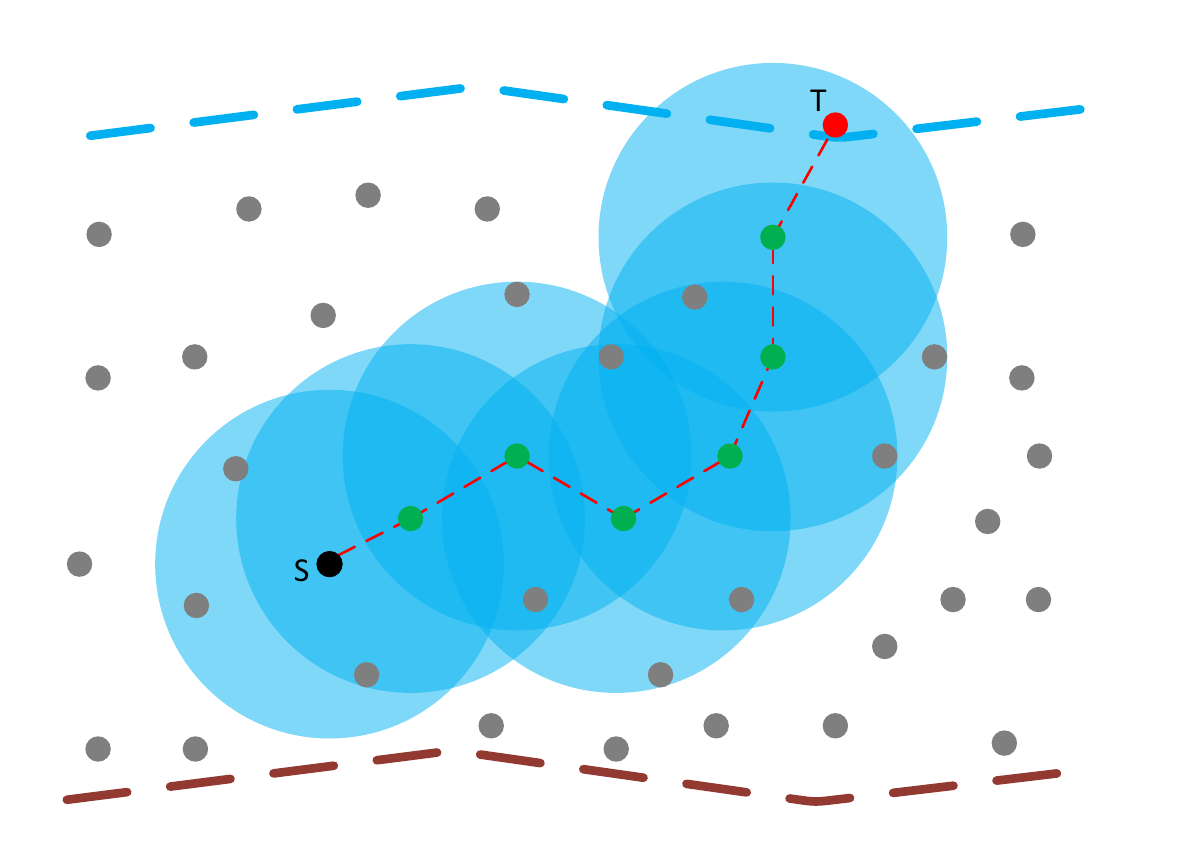}
      \caption{Distributed routing protocol (DRP).}
      \label{fig:drp}
    \end{subfigure}

    \begin{subfigure}[b]{0.35 \textwidth}
\includegraphics[width=1 \columnwidth]{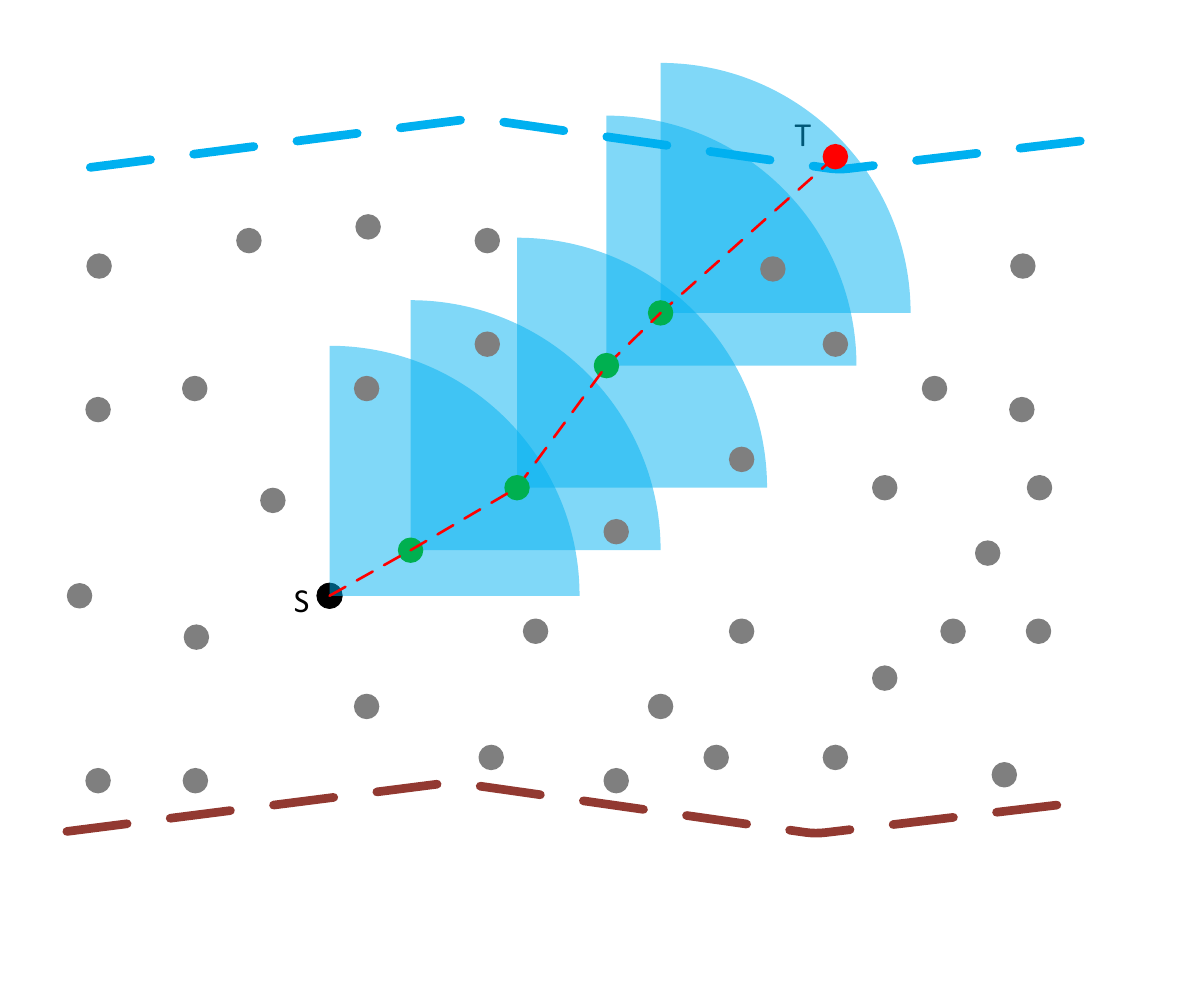}
      \caption{Sectorized routing protocol protocol (SRP).}
      \label{fig:e-drp} 
    \end{subfigure}
    \caption{End-to-end path characterization of all the three routing protocols.}
\label{fig:routing protocols}
\end{figure}

\subsection{Centralized Routing Protocol } 
Recently, a centralized routing protocol (CRP) has been proposed in \cite{Nasir} which chooses a route that provides the least end-to-end BER. In CRP, a central node collects all the neighborhood information from all the nodes in the network and create a network graph $G=(V, E)$ with $V$ set of vertices and $E$ set of weighted edges. The edges are undirected and have explicit weights, where the weight is defined as $P(e)$. Once the network graph is formed, then the central node computes all possible paths from the source node to the target node and, finally selects the best path based on the decision metric. It is also important to mention that the selected route is the one which does not visit a vertex more than once and the forwarding nodes are aware of the target's location. The decision metric for CRP is to select the least-weighted path based on Dijkstra's algorithm, in this case, the least-weighted path is the path with least end-to-end BER, as shown in Fig.~\ref{fig:crp}. In Fig.~\ref{fig:crp} the red line shows the selected route between the source and the target node while the green circles represent the forwarding nodes towards the target node. The problems of CRP include dependency on the central node, high complexity i.e., $O(N^2)$ \cite{SPP}, omnidirectional setup, and large end-to-end delay.

\subsection{ Distributed Routing Protocol}
The distributed routing protocol (DRP) is the combination of nearest neighbor problem \cite{NNP} and the traveler salesman problem \cite{TSP}, which is a special greedy version of the CRP.  Unlike the CRP, the DRP does not require a centralized node; therefore, each vertex/node has only the information of its neighbors. The edges in DRP are also undirected and are weighted as in the case of CRP, yet the route selection is based on the information known to each forwarding node.  Each forwarding node compares the edges, and the edge which has the minimum weight (least bit error rate) to its neighbors is chosen as a forwarding node until the target is found.
Fig.~\ref{fig:drp} shows a realization of the DRP in the same setup as in the CRP. The red line represents the selected route between the source and the target. It is possible in some cases that DRP may not find a closed path. In other words, the transmitted information may reach a node which is not connected to any other nodes other than previous hops (visited nodes), where additional details of the DRP is provided in Algorithm \ref{alg:dist}. The selected route based on distributed protocol would not be as good as the centralized but has a less complexity, i.e., $O(kN)$, where $k$ is the number of hops. Although DRP has less complexity compared to the CRP, yet, it passes through more number of nodes to reach the target and therefore has a large end-to-end delay. 
\begin{algorithm}
\SetAlgoLined
\KwResult{Selected path}
 UOWSN with $N$ number of nodes, and maximum transmission range $r_m$\;
 $S$ =  Source node\; 
 $T$ = Target node\;
 $k$ = Number of hops\;
 Array = Array of the BER values\;
 Path = Selected path\;
 $H$ = Current hop = $S$\;
 $H$.status = visited\;
 Array = []\;
 Path = []\;
 \For{$V = 1:N-1$}{
 	\For {$Q = 1:N-k$}{
   Array = [ Array $P(H,Q)$]
   }
  $X$ = min(Array)\;
   Path = [Path $X$]\;
   $X$.status = visited\; 
   $H = X$\;
   \If{$H == T$}{
   break\;}
 }
\caption{Distributed Routing Protocol (DRP)}\label{alg:dist}
\end{algorithm}

\subsection{Sectorized Routing Protocol (SRP)}
The proposed sectorized routing  protocol (SRP) overcomes the limitations of the aforementioned CRP and DRP.  In SRP, the decision of a forwarding node is also made based on the minimum weight (BER). However, SRP only considers the nodes which are in a quadrant towards the target node instead of considering all neighbors. This strategy is more practical for UOWSNs as the underwater optical transmission is directive and also it reduces the search space towards the target node, and thus reduces the overall complexity and delay of the protocol. In SRP,  the network is divided into four quadrants based on the location of the forwarding node and the location of the target node, in which a node/vertex compares the BER to each neighboring node within a quadrant. The source node selects the forwarding node to which the BER is minimum within the quadrant and this strategy continues until the target is found. Fig.~\ref{fig:e-drp} shows a realization of a selected route using SRP, where the red line represents the selected route. Since SRP divides the network into four quadrants for each forwarding node and selects one of them, the number of hops is much less than in the case of DRP. Therefore, the complexity of the proposed routing protocol is $O(kN)$, but in this case, $k$ is much smaller as compared to the DRP. The details of the proposed SRP is presented in Algorithm~\ref{alg:prop}.

\begin{algorithm}
\SetAlgoLined
\KwResult{Selected path}
 UOWSN with $N$ number of nodes, and maximum transmission range $r_m$\;
 $S$ =  Source node\; 
 $T$ = Target node\;
 $k$ = Number of hops\;  $U$ =  Number of nodes outside the quadrant the target (T) lies in\;
 Array = Array of the BER values\;
 Path = Selected path\;
 $H$ = Current hop = $S$\;
 $H$.status = visited\;
 Array = []\;
 Path = []\;
 \For{$V = 1:N-1$}{
 	\For {$Q = 1:N-(k+U)$}{
   Array = [ Array $P(H,Q)$]
   }
   $X$ = min(Array)\;
   Path = [Path $X$]\;
   $X$.status = visited\; 
   $H = X$\;
   \If{$H == T$}{
   break\;}
   
 }
 \caption{Sectorized Routing Protocol (SRP)}\label{alg:prop}
\end{algorithm}
\begin{figure}
\includegraphics[width=0.8\columnwidth]{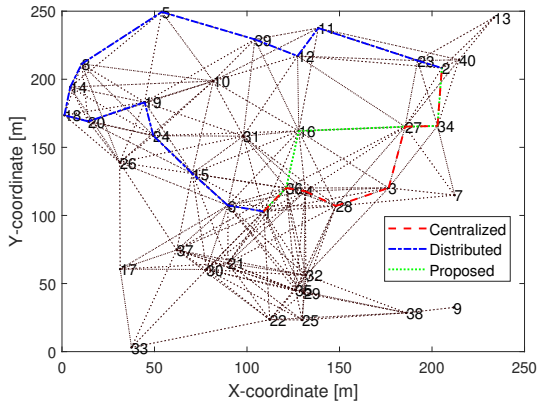}  
\caption{The selected route from source node (1) to the target node (2) for all of the three routing protocols. \label{fig:paths}}  
\end{figure}
\begin{figure*}[!htb]
\minipage{0.32 \textwidth}
\includegraphics[width=1\columnwidth]{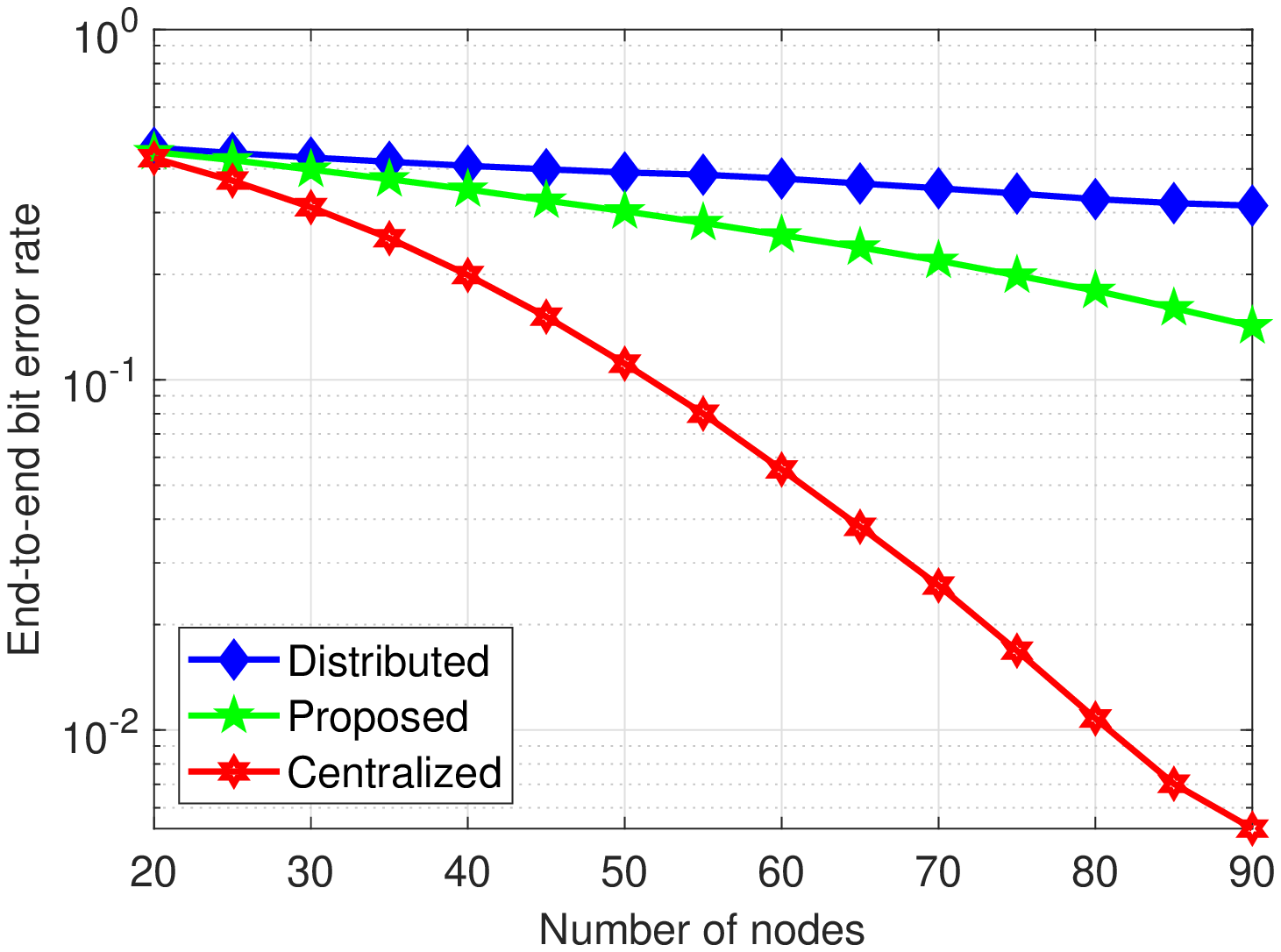}  
\caption{End-to-end BER performance.\label{fig:ber}}  
\endminipage\hfill
\minipage{0.33 \textwidth}
\includegraphics[width=1\columnwidth]{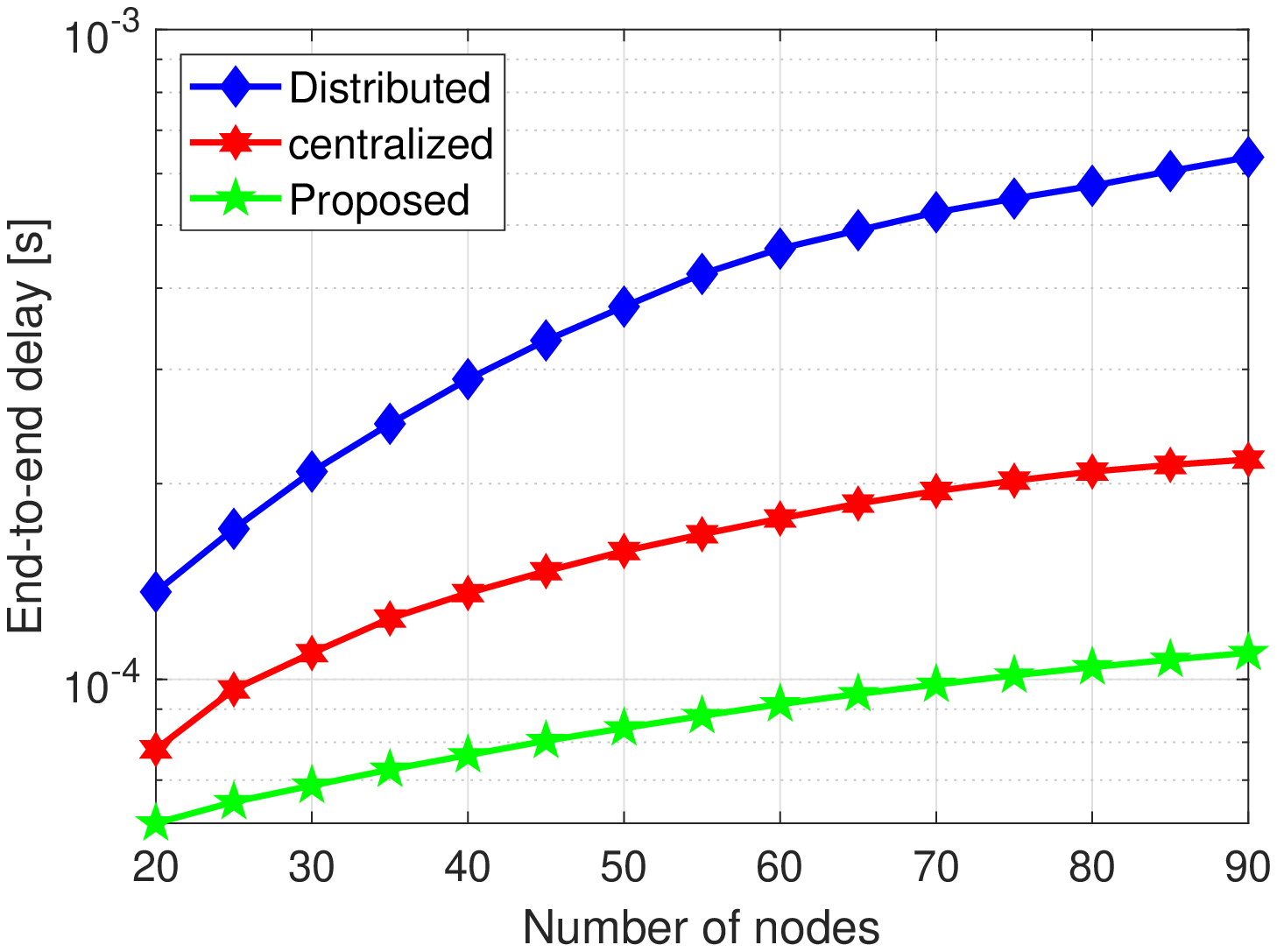}  
\caption{End-to-end delay performance.\label{fig:delay}}  
\endminipage\hfill
\minipage{0.32 \textwidth}%
\includegraphics[width=1\columnwidth]{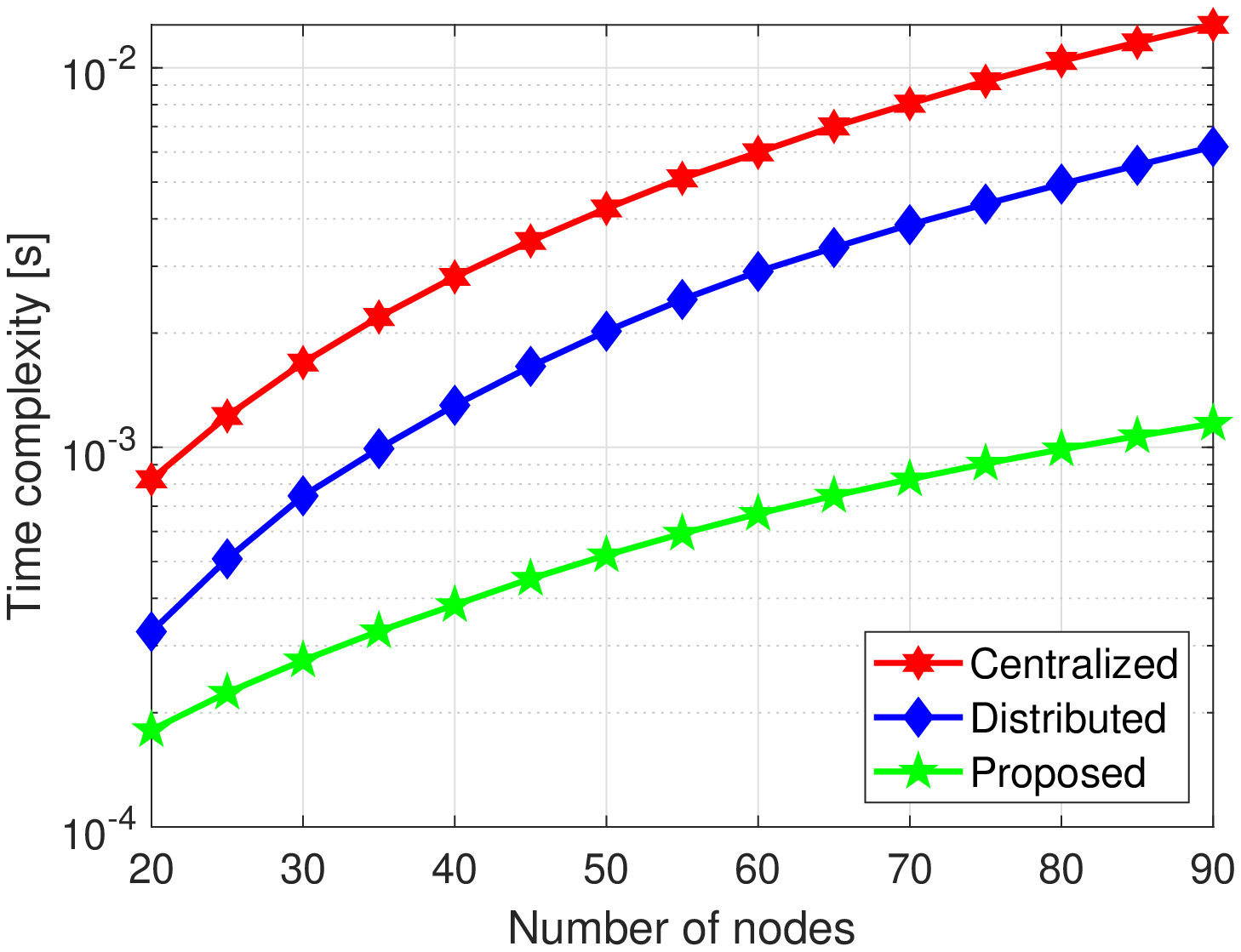}  
\caption{Time complexity performance.\label{fig:time}}  
\endminipage 
\vspace{0.3cm}
\hrule
\end{figure*}

\subsection{Numerical Results}\label{results}
In this section, we assess the performance of all the three routing protocols by providing numerical results. The performance evaluation metrics considered in this work for the three protocols are the end-to-end BER, end-to-end delay, and time complexity when the number of nodes in the network increases. The reason we are evaluating  the end-to-end BER performance is to ensure the protocols are providing a reliable communication link. Additionally, the end-to-end delay evaluation metric is a measurement to examine whether the protocols can support real-time applications or not. Finally, the time complexity evaluation metric is considered to decide which protocol has less complexity and improves the network lifetime. Fig.~\ref{fig:paths} shows a realization of the three protocols, where the source is  set to node 1 and the target is set to node 2. The total number of nodes in the network is 40 and maximum transmission range $r_m = 80~ m$. The red, blue, and green lines represent the selected route between the source and the target node based on CRP, DRP, and SRP respectively. Although CRP has good performance in terms of $P_{e2e}(e)$ but has high complexity and large end-to-end delay \cite{SPP}.
\subsubsection{Simulations Parameters}
We consider a network with a fixed source and a fixed target over 500 randomizations. The nodes are randomly distributed within an area of  $250\times250 ~m^2$. The maximum transmission range is kept to $80~ m$ to obtain a connected network. 
The data rate is kept to $ 1~Mbps$, and the distance between the source and the target is $145~ m$. The parameters listed in Table.~\ref{table:para} are mainly used throughout the simulations unless stated otherwise.

\subsubsection{End-to-End Bit Error Rate}
Fig.~\ref{fig:ber} shows the impact of increasing the number of nodes in the network on the performance of the end-to-end BER for the three routing protocols. It is clear that the CRP has a better performance than its counterparts DRP and SRP because the CRP selects the best path with least $P_{e2e}(e)$ after calculating all the possible paths. Moreover, it shows a significant decrease of $P_{e2e}(e)$ when the networks become denser. The DRP gives the worst behavior among the three, we refer this behavior to the nature of the algorithm, since it passes through too many nodes, and is possible to almost pass by every node in the network. In other words, since the DRP selects a large number of hops, it has a worse bit error rate. Even though the SRP does not provide the least $P_{e2e}(e)$, yet, it decreases with the increase in the number of nodes and has better performance than DRP. Intuitively, the selected quadrant at each forwarding node will have more nodes when the number of nodes is increased and thus reduces the end-to-end BER.

\subsubsection{End-to-End Delay}
Fig.~\ref{fig:delay} shows the end-to-end delay performance of the three routing protocols, which is dependent on the distance traveled across the network and the number of forwarding nodes in the selected path. It can be seen in Fig.~\ref{fig:delay} that the end-to-end delay increases with the increase in the number of nodes  because the number of forwarding nodes and/or the distance traveled increases. Furthermore, the DRP provides high latency,  as it travels more distance, and passes through a large number of forwarding nodes when compared to CRP. The proposed SRP provides the least delay among the three since it selects the forwarding nodes within a quadrant out of the four quadrants of the network towards the target node. Although the CRP travels short distances (to take least $P(e)$), yet the total distance traveled, and the number of forwarding nodes is higher than the proposed SRP. Also, all the three protocols have an increasing delay when the number of nodes in the network  increases, because all protocols are looking for nodes with least $P(e)$ and so, it forces the protocols to selects more number of forwarding nodes with shorter distances.

\subsubsection{Time Complexity}
Finally, we have shown the impact of an increasing number of nodes on the time complexity of the three protocols in Fig.~\ref{fig:time}. The time complexity of all the three routing protocols increases with an increase in the number of nodes which is due to the increase in the number of forwarding nodes. It is clear from Fig.~\ref{fig:time} that the proposed  SRP protocol has the least time complexity.  The  time complexity of the proposed protocol is less than the DRP  because it passes by less number of forwarding nodes (less number of candidates) to the target. Additionally, the CRP has the highest time complexity due to nature of the protocol as all the  neighbored information is sent to the centralized node and then the centralized node computes all possible routes from the source to the target and finally selects the best path.
\begin{table}
 \caption{Simulation Parameters}\label{table:para}
\begin{tabular}{ |l|l|l|l| }
 \hline
  Parameters                                 & Value    &  Parameters                   & Value \\
 \hline
  t($\lambda$) Clear ocean    $[m^{-1}]$     & $0.15$   &$r_{bg} ~[MHz]$                & $ 1$\\  
  t($\lambda$) Coastal ocean  $[m^{-1}]$     & $0.30$   &$r_{dc} ~[MHz]$                & $ 1 $\\   
  t($\lambda$) Turbid harbor  $[m^{-1}]$     & $2.19 $  &   $A ~[mm^2]$                 & $0.17$\\
  $P_{Tx}~[W]$                               & $0.1 $   &   $\theta$                    & $60^\circ$\\  
  Plane setting $[m]$                        & $250^2$  &   $R ~[Mbps] $                & $1 $ \\
  \# of realizations                         & $500$    &   $T ~[ns]$                   & $1$ \\
  Tx-Rx distance $[m]$                       & $145 $   &   $\lambda ~[nm]$             & $530 $\\
    $r_m ~[m]$                               & $80$     &   $\eta$                      & $0.9$\\  
  $\eta_T$                                   & $0.9$    &   $\eta_R$                    & $0.9$      \\           
 \hline 
\end{tabular}
\end{table}

\section{Future Research Directions}
Although some progress has been made in the area of UOWC, the results of existing studies open new avenues for future research directions which require further investigation. In the following, we highlight some of these future directions.
\begin{itemize}

\item \textbf{Channel Models:} Investigation and analysis are required for new theoretical channel models, both analytically and computationally. The new models will help in understanding the propagation of light over different types of underwater channels. It is a challenging task to find analytic and computational models because the analytic solutions are either intractable or have a high computational complexity. Therefore, there is a  dire need for tractable and accurate channel models for UOWC. Furthermore, the literature on UOWC is mainly focused on horizontal LoS configurations. However, the effect of depth, temperature, and properties of water at different layers on the performance of  UOWC  needs to be analyzed. Additionally, the NLoS  UOWC channel models have not been examined properly. Specifically, the random movement of the sea surface needs to be considered for the NLoS UOWC channel.

\item \textbf{Cross Layer Optimization:} Future studies on UOWSNs need to include the design of a cross-layer optimization framework to establish a  functional  network. Moreover, examining the adaptive techniques  for cross-layer design optimization may improve security, prevent overcrowding, maintain connectivity, and satisfy QoS requirements. 

\item  \textbf{Network Management:} One of the challenges faced by UOWC is the hardware-based system architecture, which does not allow a space for supporting new technologies such as software-defined networks (SDN). SDN is a flexible networking model which allows separating the data plan and control plan of a network for efficient network management by using an SDN controller \cite{survey}. In \cite{SoftWater}, the authors have introduced an SDN based network architecture for  UWC systems called SoftWater. However, placement of the SDN controller, traffic balancing protocols, and implementation of a wireless hypervisor are some of the future research challenges for SDN based UOWSNs.

\item  \textbf{Energy Consumption:} The sensors deployed in UOWSNs are battery-powered and thus have limited lifetime. Given the limited energy resources and the difficulties of replacing the batteries of underwater optical sensor nodes, it is crucial to develop a method for an energy self-sufficient UOWSNs to increase the lifetime of the network. One solution for limited energy resources is to empower the underwater optical sensor nodes with energy harvesting.  Research on harvesting energy from the ambient sources have been conducted on a  physical level and have shown a significant impact on the performance of the network \cite{terrestrial}. However, the methods presented in \cite{terrestrial} are not suitable for an aquatic medium.  Therefore, there is  an urgent need to develop energy harvesting schemes for UOWSNs. Also, the lifetime of the UOWSNs can be improved by introducing energy efficient routing protocols for multi-hop communication which is still an open research problem. 

\item \textbf{Deployment:} Future research directions also include studying the practical implementation of UOWSNs which includes testing and analyzing of UOWC systems in a real environment, designing of smart and compact transceivers, link management, interference management, and power budget. Additionally, a hybrid implementation which can support RF, acoustic, and optical carriers is an interesting and open research area. In fact, it is  essential to develop an intelligent environment detection algorithm to enhance the performance of a hybrid system by switching adaptively between different communication modes.  

\item \textbf{Advanced Routing Protocols:} Since most of the research efforts on UOWCs are directed towards overcoming physical layer issues, this provides a good starting point for discussion and further research on the networking directions of UOWC. Only a few works have focused on this aspect \cite{Nasir, connectivity, connectivity2, network6, network7, network8}. As the research work on networking issues is limited for UOWSNs, it opens the door for future research to develop sufficient network architecture and protocols to overcome the limited communication ranges and to increase the higher user capacity in UOWC systems. Further investigation in this area can be quite beneficial to the UOWSNs research since most of the real-time applications necessitate for efficient and reliable networking.  Future research work also needs to investigate the effect of other underwater channel impairments such as turbulence, salinity, and air-bubbles on the existing routing protocols. Additionally, future work also includes selecting a forwarding node not only based on the minimum BER and to be in a quadrant towards the target node but also to consider energy, reliability, and beam-width of the forwarding node. 

\end{itemize}

\section{Final Remarks}\label{conc}
In this paper, we have proposed distributed routing protocols for UOWSNs with low complexity and low end-to-end delay. We have considered a LoS optical wireless channel in clear ocean water with different densities of the sensor nodes. Although the end-to-end BER performance of the proposed distributed routing protocols is less than the centralized routing protocol, it reduces the complexity of the network by distributing the load over the network. Also, the proposed distributed routing protocols reduces the end-to-end transmission delay. Numerical results have been provided which validates the performance of the proposed routing protocols. The proposed sectorized routing protocol has less complexity and low end-to-end delay in the presence of  absorption, scattering, and geometrical losses.  Finally, we have outlined  some future research directions in UOWSNs research.

\bibliographystyle{../bib/IEEEtran}
\bibliography{../bib/IEEEabrv,../bib/nasir_ref}

\end{document}